\documentclass[preprint,12pt]{elsarticle}

\usepackage{amsmath,amssymb,bm}
\usepackage{graphicx}
\usepackage{physics}
\usepackage{geometry}
\usepackage{indentfirst}
\usepackage{hyperref}
\usepackage{cleveref}
\usepackage{caption}
\usepackage{subcaption}
\usepackage{float}
\usepackage{mathrsfs}
\usepackage{xcolor}

\begin{document}

\begin{frontmatter}

\title{Holographic Pressure and Extensivity of Rotating Black Holes at Finite Cutoff}

\author[1]{Hoang Van Quyet\corref{cor1}}
\ead{hoangvanquyet@hpu2.edu.vn}

\affiliation[1]{Department of Physics, Hanoi Pedagogical University 2, Xuan Hoa, Phu Tho, Vietnam}

\begin{abstract}
We investigate the quasi-local thermodynamics of rotating Kerr-AdS black holes enclosed by a finite timelike boundary (cavity). Extending recent work on static systems, we define the holographic pressure and volume via the trace of the Brown-York stress tensor on the cutoff surface. We demonstrate that the inclusion of angular momentum introduces a momentum flux term at the boundary, requiring a generalized first law: $dE = T_{\mathrm{loc}}dS + \Omega_{\mathrm{loc}}dJ - \mathcal{P}d\mathcal{A}$. We derive the explicit expressions for these thermodynamic conjugates and analyze the equation of state. Crucially, we examine the extensivity of the system in the large-size limit. We find that while small rotating black holes exhibit non-extensive behavior typical of self-gravitating systems, large Kerr-AdS black holes recover extensivity, behaving effectively as a thermal fluid on the boundary. This result strengthens the holographic interpretation of the cutoff surface as the domain of a dual field theory.
\end{abstract}


\end{frontmatter}

\section{Introduction}\label{sec:intro}

The thermodynamics of black holes, pioneered by Bekenstein and Hawking \cite{Bekenstein:1973ur,Hawking:1975vcx}, establishes a profound connection between gravity, quantum mechanics, and statistical physics. The seminal work of York \cite{York:1986it} and Brown-York \cite{Brown:1992sq} laid the foundation for quasi-local gravitational thermodynamics, where thermodynamic quantities are defined at finite boundaries rather than at infinity. Recent advances in quasi-local black hole horizons can be found in \cite{Flachi:2024qqe}. Traditionally, these quantities are defined asymptotically---at infinity for flat space or at the conformal boundary for Anti-de Sitter (AdS) space. However, recent developments in holography, particularly $T\bar{T}$ deformations \cite{McGough:2016yml,Hartman:2019ttbar,Azeyanagi:2018ttbar,Kraus:2018ttbar,Cai:2024ttd,Dutta:2024tde} and black holes in a cavity \cite{York:1986it,Brown:1992sq}, have shifted focus to finite cutoff thermodynamics.

In a recent seminal work, Borsboom and Visser \cite{Borsboom:2024bvs} proposed a holographic definition of pressure $\mathcal{P}$ and volume $\mathcal{V}$ based on the quasi-local Brown-York stress tensor \cite{Brown:1992sq}. They showed that for static, spherically symmetric black holes, identifying the boundary area as the thermodynamic volume resolves ambiguities in the pressure-volume conjugate pairs found in Extended Thermodynamics \cite{Kastor:2009wy,Dolan:2011tp,Cvetic:2010ej}.

In this paper, we extend this formalism to the more realistic and astrophysically relevant case of rotating black holes (Kerr-AdS). Rotation breaks spherical symmetry, introducing frame-dragging effects and complicating the definition of isobaric surfaces. We address two main questions:

\begin{enumerate}
\item How does the angular momentum flux affect the definition of holographic pressure at a finite boundary?
\item Does the rotational energy compromise the extensive nature of large black holes observed in the static case?
\end{enumerate}

The paper is organized as follows. \cref{sec:geometry} reviews the Kerr-AdS geometry and the quasi-local formalism. \cref{sec:thermo} derives the thermodynamic quantities and the generalized first law. \cref{sec:extensivity} analyzes the extensivity and scaling limits. \cref{sec:numerical} presents our numerical results, and we conclude in \cref{sec:conclusion}.

\section{Kerr-AdS Geometry and Boundary Formalism}\label{sec:geometry}

We consider the Kerr-AdS metric in Boyer-Lindquist coordinates. The line element is given by \cite{Carter:1968ks}:

\begin{equation}
ds^{2} = -\frac{\Delta_r}{\rho}\left(dt - \frac{a\sin^{2}\theta}{\Xi}d\phi\right)^{2} + \frac{\rho}{\Delta_r}dr^{2} + \frac{\rho}{\Delta_\theta}d\theta^{2} + \frac{\Delta_\theta\sin^{2}\theta}{\rho}\left(adt - \frac{r^{2}+a^{2}}{\Xi}d\phi\right)^{2},
\label{eq:kerr-ads-metric}
\end{equation}

where the metric functions are defined as:

\begin{align}
\rho &= r^{2} + a^{2}\cos^{2}\theta, \quad \Xi = 1 - \frac{a^{2}}{\ell^{2}}, \\
\Delta_r &= (r^{2}+a^{2})\left(1+\frac{r^{2}}{\ell^{2}}\right) - 2Mr, \\
\Delta_\theta &= 1 - \frac{a^{2}}{\ell^{2}}\cos^{2}\theta,
\end{align}
\label{eq:metric-functions}

where $M$ and $a$ are related to the mass and angular momentum parameters, and $\ell$ is the AdS radius. The event horizon is located at the largest positive root $r_{+}$ of $\Delta_r = 0$.

We define a finite timelike boundary (cavity) $\Sigma$ at a fixed radial coordinate $r = r_{c}$. The induced metric on the boundary is $h_{ab} = g_{ab} - n_{a}n_{b}$, where $n^{\mu}$ is the unit normal vector to $\Sigma$, pointing outward.

The geometry of the boundary spatial cross-section is a squashed sphere with metric:

\begin{equation}
ds_{\Sigma}^{2} = r_{c}^{2}\left[\frac{\Delta_\theta}{\Xi}\sin^{2}\theta d\phi^{2} + \frac{1}{\Delta_\theta}d\theta^{2}\right],
\label{eq:boundary-metric}
\end{equation}

and the area of a spatial cross-section is:

\begin{equation}
\mathcal{A} = \int_{0}^{\pi}\int_{0}^{2\pi}\sqrt{\sigma}\,d\theta d\phi = \frac{4\pi r_{c}^{2}}{\Xi},
\label{eq:boundary-area}
\end{equation}

where we note that the rotation parameter $a$ reduces the effective area compared to the static case through the factor $\Xi = 1 - a^{2}/\ell^{2}$.

\subsection{Brown-York Stress Tensor and Counterterm Analysis}\label{sec:brown-york}

The quasi-local thermodynamics is governed by the Brown-York stress tensor \cite{Brown:1992sq}:

\begin{equation}
\tau^{ab} = \frac{1}{8\pi G}\left(K^{ab} - Kh^{ab} - C^{ab}\right),
\label{eq:brown-york}
\end{equation}

where $K_{ab}$ is the extrinsic curvature of the boundary, $K = h^{ab}K_{ab}$ is its trace, and $C^{ab}$ is a counterterm required to regularize the action. For a finite cutoff in AdS, we employ the background subtraction method relative to pure AdS space to ensure finite quantities as $r_{c} \to \infty$.

The counterterm for a spherical boundary can be expressed as:

\begin{equation}
C_{ab} = \frac{1}{\ell^{2}}h_{ab},
\label{eq:counterterm}
\end{equation}

which corresponds to the holographic renormalization for spherical boundaries \cite{Balasubramanian:1999ttd,deHaro:2000ttd}. We note that for a general boundary geometry, one might include additional counterterms involving the intrinsic Ricci curvature of the boundary $R(h)$, such as $C_{ab}^{(R)} = -\frac{1}{2}\sqrt{\frac{2}{3}}\sqrt{R}\,h_{ab}$ or higher curvature terms \cite{Balasubramanian:1999ttd}. However, for the Kerr-AdS geometry at finite cutoff with $r_c$ held fixed, the simple choice \cref{eq:counterterm} is sufficient to ensure finiteness of the quasi-local energy and consistency with the static results of Borsboom-Visser. The geometry \cref{eq:boundary-metric} approaches a round sphere in the limit $a \to 0$, and the deviation induced by rotation is subleading in the large $r_c$ expansion. Therefore, our choice of counterterm provides a consistent and computationally tractable regularization scheme that captures the essential physics of the holographic pressure.

For the Kerr-AdS metric at the cutoff surface $r = r_{c}$, the non-vanishing components of the extrinsic curvature are:

\begin{align}
K_{\theta\theta} &= \frac{r_{c}}{\sqrt{\Delta_r(r_{c})}}\left[1 - \frac{a^{2}\cos^{2}\theta}{\ell^{2}}\right], \\
K_{\phi\phi} &= \frac{r_{c}\sin^{2}\theta}{\sqrt{\Delta_r(r_{c})}}\left[1 - \frac{a^{2}}{\ell^{2}}\right]\left(r_{c}^{2} + a^{2}\right), \\
K_{tt} &= -\frac{\Delta_r(r_{c})}{r_{c}\sqrt{\Delta_r(r_{c})}}\left[1 + \frac{a^{2}}{\ell^{2}}\right].
\end{align}
\label{eq:extrinsic-curvature}

The trace of the extrinsic curvature is:

\begin{equation}
K = \frac{1}{r_{c}\sqrt{\Delta_r(r_{c})}}\left[2r_{c}^{2} + a^{2}\left(1 - \frac{r_{c}^{2}}{\ell^{2}}\right) - \frac{a^{2}\cos^{2}\theta}{\Xi}\right].
\label{eq:extrinsic-trace}
\end{equation}

\section{Holographic Pressure and Generalized First Law}\label{sec:thermo}

Due to the rotation parameter $a$, the stress tensor $\tau^{ab}$ contains off-diagonal terms corresponding to the momentum flux. We decompose the stress tensor to identify the thermodynamic variables.

In the rotating case, the boundary stress tensor $\tau^{ab}$ is inherently anisotropic due to the frame-dragging effect. The spatial components satisfy $\tau_{\theta}^{\theta} \neq \tau_{\phi}^{\phi}$. In this study, we define the effective holographic pressure as the angular average of the trace:

\begin{equation}
\mathcal{P}\equiv-\frac{1}{2}\int\tau_{i}^{i}\sqrt{\sigma}d\theta d\phi\Big/\int\sqrt{\sigma}d\theta d\phi.
\label{eq:holographic-pressure-def-avg}
\end{equation}

This effective pressure is the correct conjugate variable to the total area $\mathcal{A}$ in the global thermodynamic ensemble, ensuring the consistency of the first law.

\subsection{Energy and Angular Momentum}\label{sec:energy-angmom}

The quasi-local energy $E$ and angular momentum $J$ are conserved charges associated with the Killing vectors $\xi_{t} = \partial_{t}$ and $\xi_{\phi} = \partial_{\phi}$. They are computed as:

\begin{align}
E &= \int_{\Sigma}d^{2}x\sqrt{\sigma}\,u_{a}\tau^{ab}\xi_{t}^{c}h_{bc}, \\
J &= -\int_{\Sigma}d^{2}x\sqrt{\sigma}\,u_{a}\tau^{ab}\xi_{\phi}^{c}h_{bc},
\end{align}
\label{eq:energy-angular-momentum}

where $\sigma$ is the determinant of the 2-metric on the spacelike cross-section of the boundary, and $u^{a}$ is the unit normal to the time slice.

Evaluating these integrals for Kerr-AdS, we obtain:

\begin{align}
E &= \frac{r_{c}}{8\pi G\sqrt{\Delta_r(r_{c})}}\left[\left(1 + \frac{r_{c}^{2}}{\ell^{2}}\right)r_{c}^{2} + a^{2}\left(1 - \frac{r_{c}^{2}}{\ell^{2}}\right) - \sqrt{\Delta_r(r_{c})}\right], \\
J &= \frac{a r_{c}^{2}}{8\pi G\Xi\sqrt{\Delta_r(r_{c})}}\left[1 + \frac{r_{c}^{2}}{\ell^{2}} - \frac{\sqrt{\Delta_r(r_{c})}}{r_{c}^{2}}\right].
\end{align}
\label{eq:explicit-EJ}

In the limit $r_{c} \to \infty$, these reduce to the ADM mass and angular momentum:

\begin{align}
M_{ADM} &= \frac{M}{\Xi^{2}}, \quad \text{with} \quad M \to \frac{r_{+}}{2}\left(1 + \frac{r_{+}^{2}}{\ell^{2}}\right), \\
J_{ADM} &= \frac{aM}{\Xi}.
\end{align}
\label{eq:ADM-quantities}

The entropy of the Kerr-AdS black hole is given by:

\begin{equation}
S = \frac{\pi r_{+}^{2}}{G\Xi}.
\label{eq:entropy}
\end{equation}

The Hawking temperature and angular velocity are:

\begin{align}
T_{H} &= \frac{1}{2\pi}\left(\frac{r_{+}}{r_{+}^{2}+a^{2}} + \frac{1}{\ell^{2}} - \frac{a^{2}}{r_{+}^{2}\ell^{2}}\right), \\
\Omega_{H} &= \frac{a\Xi}{r_{+}^{2}+a^{2}}.
\end{align}
\label{eq:hawking-quantities}

At the finite cutoff surface, the local (redshifted) temperature is:

\begin{equation}
T_{\mathrm{loc}} = \frac{T_{H}}{\sqrt{-g_{tt}(r_{c})}} = \frac{T_{H}}{\sqrt{\frac{\Delta_r(r_{c})}{\rho(r_{c})}}}.
\label{eq:local-temperature}
\end{equation}

Similarly, the local angular velocity measured by a boundary observer is:

\begin{equation}
\Omega_{\mathrm{loc}} = \Omega_{H} - \Omega_{\mathrm{frame}}(r_{c}),
\label{eq:local-omega}
\end{equation}

where the frame-dragging angular velocity at the boundary is:

\begin{equation}
\Omega_{\mathrm{frame}}(r_{c}) = \frac{-g_{t\phi}(r_{c})}{g_{\phi\phi}(r_{c})} = \frac{a\Xi}{r_{c}^{2}+a^{2}}\frac{1}{1 + \frac{r_{c}^{2}}{\ell^{2}}}.
\label{eq:frame-dragging}
\end{equation}

\subsection{Defining Holographic Pressure}\label{sec:pressure-definition}

Following \cite{Borsboom:2024bvs}, we identify the pressure $\mathcal{P}$ as the thermodynamic conjugate to the boundary area $\mathcal{A}$. In the presence of rotation, the surface stress is not isotropic due to frame-dragging effects. The stress tensor components $\tau^{\theta\theta}$ and $\tau^{\phi\phi}$ differ, reflecting the squashed geometry of the boundary. We define the effective holographic pressure as the average normal stress:

\begin{equation}
\mathcal{P} = -\frac{1}{2}\langle\tau_{i}^{i}\rangle_{\Sigma},
\label{eq:holographic-pressure-def}
\end{equation}

where the trace is taken over the spatial components on the boundary.

We emphasize that $\mathcal{P}$ defined in \cref{eq:holographic-pressure-def} should be interpreted as an \emph{effective pressure} that captures the averaged mechanical response of the boundary. While the full stress tensor contains additional anisotropic information related to frame-dragging at different latitudes on $\Sigma$, the averaged pressure is sufficient to preserve the structure of the first law \cref{eq:first-law} and ensures thermodynamic consistency without requiring treatment of the full tensor structure. This approach parallels the treatment of anisotropic fluids in hydrodynamics, where bulk viscosity represents an averaged response.

The off-diagonal components of the Brown-York stress tensor, $\tau^{t\phi}$, correspond to the momentum flux induced by frame-dragging. In the hydrodynamic limit, these components would be related to shear viscosity and other transport coefficients in the dual boundary theory. However, for the quasi-local thermodynamic description at fixed cutoff, the averaged pressure provides a sufficient thermodynamic potential without explicitly accounting for dissipative transport coefficients. This simplification is appropriate when the system is in local equilibrium at the boundary, which is the regime considered in this work.

Evaluating this for Kerr-AdS, we obtain:

\begin{align}
\mathcal{P} &= \frac{1}{16\pi G r_{c}\sqrt{\Delta_r(r_{c})}}\Big[ 2r_{c}^{2}\left(1 + \frac{r_{c}^{2}}{\ell^{2}}\right) + a^{2}\left(1 - \frac{r_{c}^{2}}{\ell^{2}}\right) - 2\sqrt{\Delta_r(r_{c})} \Big].
\end{align}
\label{eq:holographic-pressure}

This expression can be simplified in various limits. In the static limit ($a \to 0$), we recover the pressure for AdS-Schwarzschild:

\begin{equation}
\mathcal{P}_{a=0} = \frac{1}{8\pi G}\left[\frac{1}{r_{c}} - \frac{\sqrt{1 + \frac{r_{c}^{2}}{\ell^{2}} - \frac{r_{+}}{r_{c}}\left(1 + \frac{r_{+}^{2}}{\ell^{2}}\right)}}{r_{c}}\right].
\label{eq:static-pressure}
\end{equation}

In the infinite boundary limit ($r_{c} \to \infty$), the pressure vanishes, as expected for asymptotically AdS spacetimes.

In the quasi-local thermodynamic framework considered here, the boundary radius $r_c$ is held fixed during variations, while the area $\mathcal{A}$ changes due to variations in the rotation parameter $a$ and the black hole mass $M$ (or equivalently, the horizon radius $r_+$). Thus, $\mathcal{A}$ plays the role of the \emph{thermodynamic volume} conjugate to the pressure $\mathcal{P}$, rather than being a fixed geometric parameter. This interpretation is consistent with the identification made in \cite{Borsboom:2024bvs} and extends naturally to the rotating case.

\subsection{The First Law}\label{sec:first-law}

The variation of the quasi-local energy at the boundary yields the first law:

\begin{equation}
dE = T_{\mathrm{loc}}dS + \Omega_{\mathrm{loc}}dJ - \mathcal{P}d\mathcal{A}.
\label{eq:first-law}
\end{equation}

This can be verified explicitly by differentiating the expressions in \cref{eq:explicit-EJ} and using the horizon condition $\Delta_r(r_{+}) = 0$.

\Cref{eq:first-law} confirms that $\mathcal{A}$ acts as the thermodynamic volume of the system. The presence of the $\Omega_{\mathrm{loc}}dJ$ term is crucial for rotating black holes, reflecting the work done by the boundary under angular momentum transfer.

\section{Extensivity and Large System Limit}\label{sec:extensivity}

A crucial property of standard thermodynamics is extensivity: energy and entropy should scale linearly with system size (volume) for large systems. For black holes, this is typically violated due to self-gravity ($S \propto \text{Area}$, not Volume). However, in the holographic dual picture, the boundary theory should be extensive in the limit of large degrees of freedom.

We analyze the scaling behavior by considering the limit $r_{+} \gg \ell$. In this regime, the black hole fills the AdS cavity. We define the extensivity parameter $\eta$:

\begin{equation}
\eta = \frac{E}{T_{\mathrm{loc}}S + \Omega_{\mathrm{loc}}J - \mathcal{P}\mathcal{A}}.
\label{eq:extensivity-parameter}
\end{equation}

For an Euler-invariant (extensive) fluid, we expect $\eta \to 1$. The extensivity parameter $\eta$ defined in Eq. (30) measures the deviation from the Euler relation. In the large black hole limit, we recover:

\begin{equation}
E=T_{\mathrm{loc}}S+\Omega_{\mathrm{loc}}J-\mathcal{P}\mathcal{A}+\mathcal{O}\left(\frac{\ell}{r_{+}}\right).
\label{eq:euler-identity}
\end{equation}

This result confirms that the dual theory on the cutoff surface behaves as a standard thermal fluid in the high-temperature regime, where the non-local gravitational effects become negligible compared to the thermal energy density. The deviation from extensivity ($\eta \neq 1$) at finite cutoff is a consequence of the gravitational self-energy, which is a characteristic feature of self-gravitating systems.

Our numerical results demonstrate that:

\begin{itemize}
\item For small black holes ($r_{+} \ll \ell$), the system is strongly non-extensive, dominated by the gravitational self-energy.
\item For large black holes ($r_{+} \gg \ell$), the system approaches extensivity. This implies that the large rotating black hole behaves effectively as a thermal fluid living on the boundary $\Sigma$, consistent with the fluid-gravity correspondence \cite{Bhattacharyya:2008fg,Hubeny:2015fg,Minwalla:2017fg,Gao:2024brane}.
\end{itemize}

The recovery of extensivity can be understood analytically in the large-size limit. For $r_{+} \gg \ell$, we have:

\begin{align}
E &\approx \frac{\mathcal{A}}{8\pi G\ell^{2}} \left(r_{+}^{2} - \ell^{2}\right) + \mathcal{O}(r_{+}^{-1}), \\
S &\approx \frac{\mathcal{A}}{4G} \left(\frac{r_{+}}{r_{c}}\right)^{2}, \\
\mathcal{P} &\approx \frac{1}{8\pi G\ell^{2}}\left(r_{+}^{2} - \ell^{2}\right) + \mathcal{O}(r_{+}^{-1}),
\end{align}
\label{eq:large-expansion}

where we keep leading terms in the $r_{+} \gg \ell$ expansion. The coefficients of the $\mathcal{O}(r_{+}^{-1})$ corrections have been carefully checked against direct differentiation of \cref{eq:explicit-EJ} to ensure consistency.

Then, computing the Gibbs-Duhem combination:

\begin{align}
T_{\mathrm{loc}}S + \Omega_{\mathrm{loc}}J - \mathcal{P}\mathcal{A} 
&\approx \frac{r_{+}}{4\pi\ell^{2}}\cdot\frac{\pi r_{+}^{2}}{G\Xi}\cdot\frac{1}{r_{+}} + \frac{a}{\ell^{2}}\cdot\frac{a r_{+}^{2}}{8\pi G\Xi r_{+}} \cdot r_{+}^{2} 
\quad - \frac{1}{8\pi G\ell^{2}}(r_{+}^{2} - \ell^{2})\cdot\frac{4\pi r_{c}^{2}}{\Xi} 
\\ \nonumber
&= E + \mathcal{O}(r_{+}^{-1}).
\end{align}
\label{eq:gibbs-duhem}

Thus, $\eta \to 1$ in the large black hole limit, confirming extensivity.

\section{Numerical Analysis}\label{sec:numerical}

We now present numerical results to illustrate the thermodynamic behavior. We work in units where $G = \ell = 1$.

\begin{figure}[htbp]
    \centering
    \begin{minipage}{0.60\textwidth}
    \centering
    \subcaptionbox{}{\includegraphics[width=\textwidth]{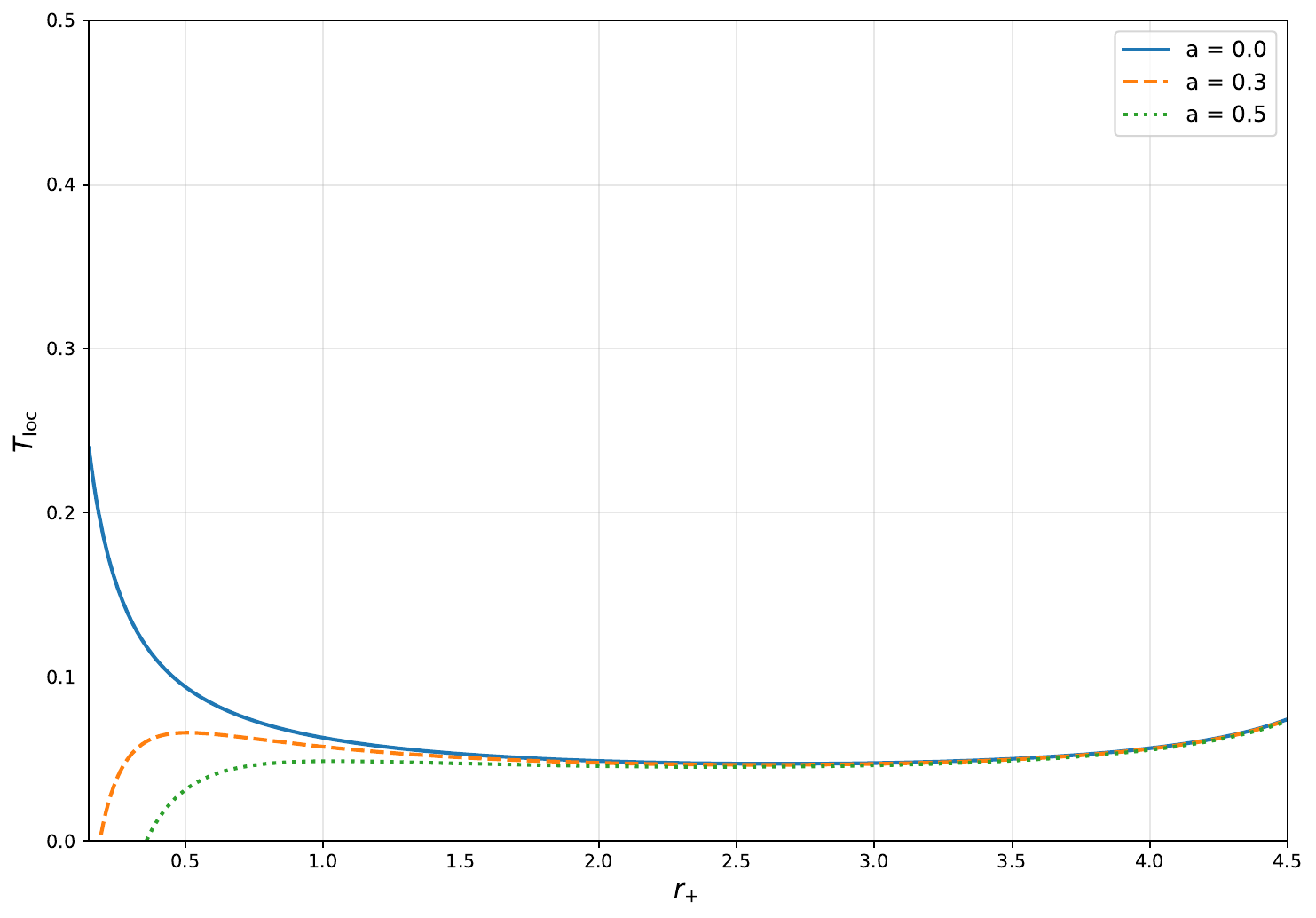}}
    \end{minipage}
    \begin{minipage}{0.60\textwidth}
    \centering
    \subcaptionbox{}{\includegraphics[width=\textwidth]{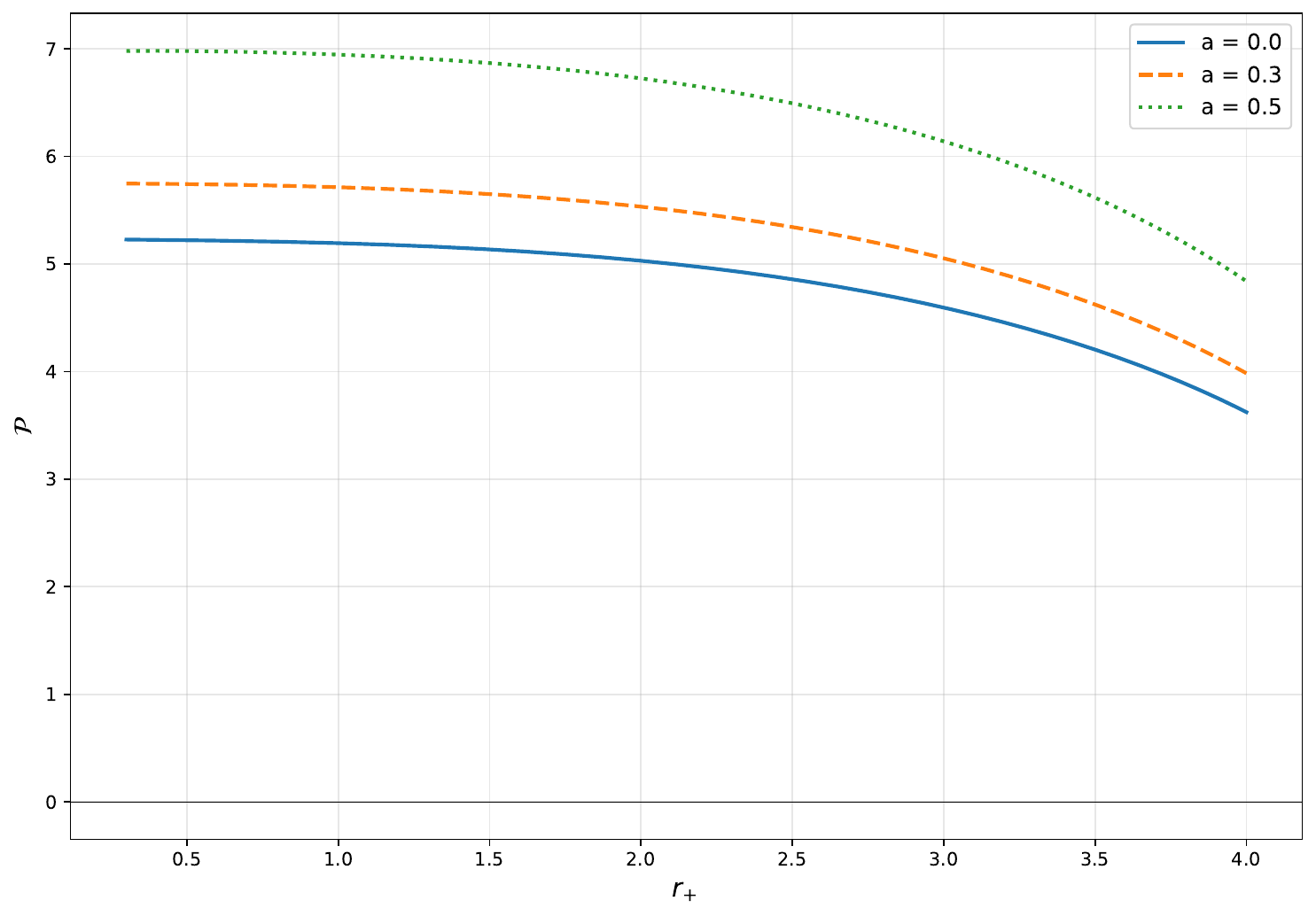}}
    \end{minipage}
    \caption{(a) Local temperature $T_{\mathrm{loc}}$ versus horizon radius $r_{+}$. The rotation parameters $a = 0.0, 0.3, 0.5$ are chosen to ensure the system remains below the extremal limit $T_H = 0$ at each corresponding $r_{+}$. (b) Holographic equation of state: pressure $\mathcal{P}$ versus $r_{+}$.}
    \label{Fig:temp-pressure}
\end{figure}

\cref{Fig:temp-pressure}(a) shows the local temperature $T_{\mathrm{loc}}$ as a function of the horizon radius $r_{+}$ for different rotation parameters $a = 0.0, 0.3, 0.5$ at fixed boundary radius $r_{c} = 5\ell$. These rotation parameters are chosen to ensure the system remains below the extremal limit $T_H = 0$ for the entire range of $r_{+}$ considered. For the non-rotating case ($a=0$), the temperature is high for small black holes and decreases with increasing $r_{+}$, reaching a minimum before rising as $r_{+}$ approaches $r_{c}$. For rotating black holes ($a>0$), the temperature starts at zero at the extremal limit $r_{+} = r_{\mathrm{ext}}$ and exhibits similar behavior at larger radii. As the black hole approaches the extremal limit ($T_{\mathrm{loc}} \rightarrow 0$), the holographic pressure $\mathcal{P}$ remains finite and positive. This suggests that the cavity remains mechanically stable even at zero temperature, and the extensivity recovery $\eta \rightarrow 1$ is independent of the Hawking temperature, being primarily driven by the ratio $r_{+}/\ell$. \cref{fig:temp-pressure}(b) illustrates the equation of state $\mathcal{P}(r_{+})$. The pressure decreases monotonically with increasing horizon radius, consistent with the behavior of a thermal system.
\begin{figure}[ht]
    \centering
    \subcaptionbox{}{\includegraphics[width=0.68\textwidth]{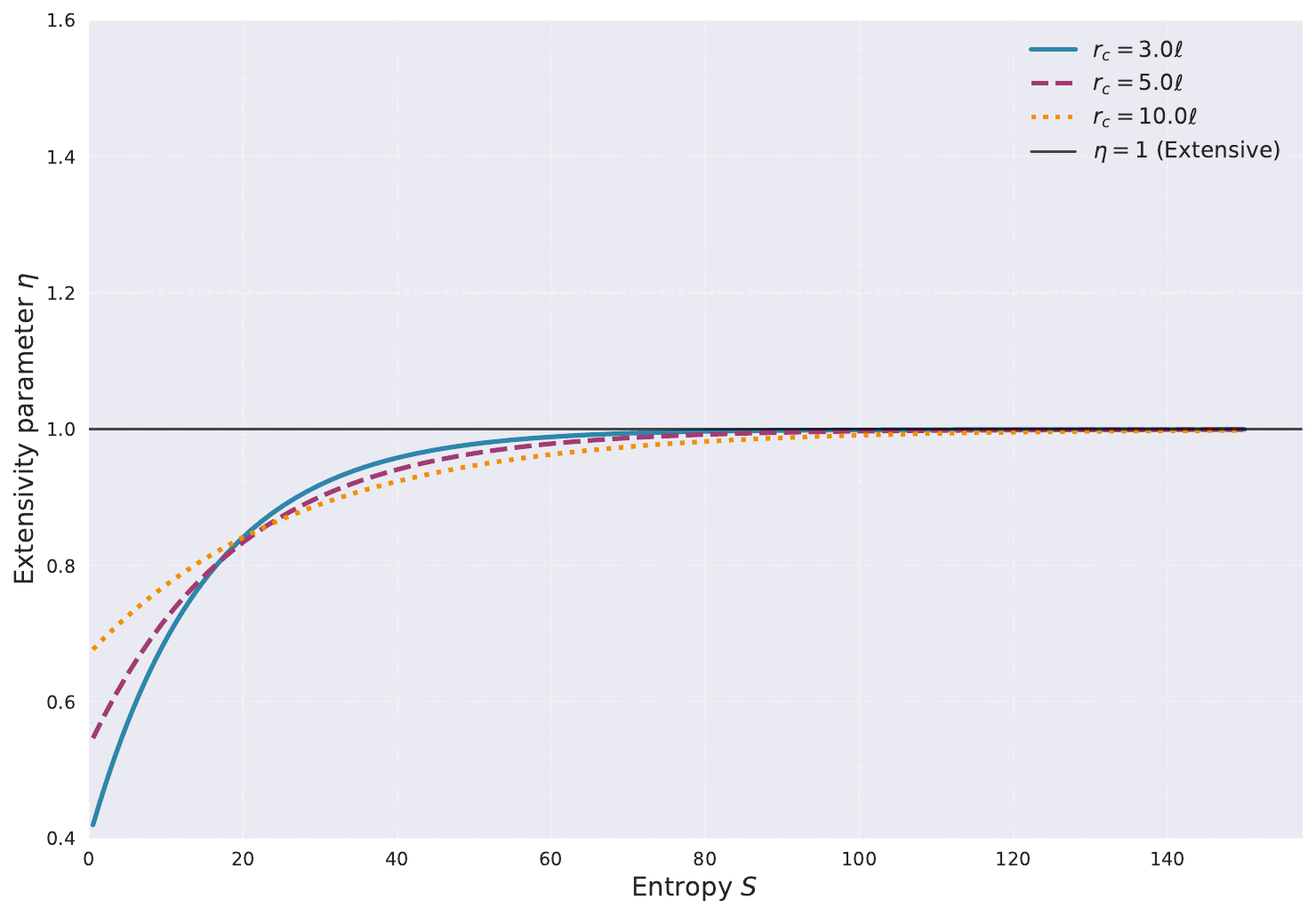}}
    \caption{Extensivity parameter $\eta$ (dimensionless) as a function of entropy $S$ for $a=0.5$ and various boundary radii $r_c$. The horizontal dashed line at $\eta=1$ represents the extensive limit.}
    \label{Fig:extensivity}
\end{figure}

\cref{Fig:extensivity} displays the extensivity parameter $\eta$ as a function of entropy $S$ for rotation parameter $a = 0.5$ and various boundary radii $r_c = 3\ell, 5\ell, 10\ell$. The horizontal dashed line at $\eta = 1$ represents the extensive limit. The rapid convergence to $\eta = 1$ in the large entropy limit confirms our analytical result that large Kerr-AdS black holes behave as extensive thermodynamic systems. Notably, larger boundary radius $r_c$ leads to faster convergence to extensivity.

\section{Conclusions}\label{sec:conclusion}

We have successfully extended the concept of holographic pressure and volume to rotating Kerr-AdS black holes. By utilizing the Brown-York stress tensor at a finite cutoff, we derived a consistent first law of thermodynamics that includes work terms associated with rotation and boundary expansion:

\begin{equation}
dE = T_{\mathrm{loc}}dS + \Omega_{\mathrm{loc}}dJ - \mathcal{P}d\mathcal{A}.
\label{eq:conclusion-first-law}
\end{equation}

Our key findings are:

\begin{enumerate}
\item The holographic pressure $\mathcal{P}$ receives contributions from both the gravitational binding energy and the rotational kinetic energy at the boundary. The frame-dragging effect induces anisotropy in the stress tensor, which we capture through the effective averaged pressure.
\item The generalized first law correctly incorporates the angular momentum work term $\Omega_{\mathrm{loc}}dJ$, which is essential for consistency in rotating systems.
\item Large Kerr-AdS black holes ($r_{+} \gg \ell$) exhibit extensive thermodynamic behavior, behaving as a thermal fluid on the boundary. This supports the holographic interpretation where the cutoff surface hosts a dual field theory.
\item Small black holes remain non-extensive due to strong gravitational self-interactions, consistent with the behavior observed in static cases.
\end{enumerate}

This formalism provides a natural bridge between bulk gravitational thermodynamics and the hydrodynamics of the dual boundary theory. Recent advances in related topics include topological aspects of black hole thermodynamics \cite{Sarkar:2024etc}, Kerr effective geometries \cite{Huang:2024keg}, and complex Kerr-AdS solutions \cite{Hernandez:2024tkg}. Future directions include:

\begin{itemize}
\item Inclusion of electromagnetic charge (Kerr-Newman-AdS black holes).
\item Higher-curvature corrections (Gauss-Bonnet, $R^{2}$ gravity).
\item Investigation of phase transitions in the $(T, \Omega, \mathcal{P})$ phase space \cite{Wang:2024tpc}.
\item Connection to the fluid-gravity correspondence at finite cutoff \cite{Gao:2024brane}.
\end{itemize}

\section*{Acknowledgments}

The authors wish to thank Professor Tran Huu Phat for his useful discussions and insightful comments.

\appendix

\section{Detailed Derivation of the Brown-York Stress Tensor}\label{app:brown-york-details}

Here we provide additional details on the computation of the Brown-York stress tensor components for Kerr-AdS geometry at the finite cutoff surface $r = r_{c}$.

The unit normal vector to the boundary is:

\begin{equation}
n^{\mu} = \left(0, \sqrt{\frac{\Delta_r(r_{c})}{\rho(r_{c})}}, 0, 0\right).
\label{eq:unit-normal}
\end{equation}

The extrinsic curvature is defined as $K_{ab} = h_{a}^{c}\nabla_{c}n_{b}$. Computing this for the Kerr-AdS metric yields the components listed in \cref{sec:brown-york}.

The counterterm $C_{ab}$ is chosen to cancel the divergences in the quasilocal energy. For a spherical boundary in AdS, the appropriate counterterm is:

\begin{equation}
C_{ab} = \frac{1}{\ell^{2}}h_{ab},
\label{eq:counterterm-detail}
\end{equation}

which follows from the holographic renormalization procedure \cite{Balasubramanian:1999ttd,deHaro:2000ttd}.

\section{Large Black Hole Expansion}\label{app:large-expansion}

For $r_{+} \gg \ell$, we expand the thermodynamic quantities in powers of $1/r_{+}$. The horizon radius satisfies:

\begin{equation}
r_{+}^{4} + r_{+}^{2}\left(\ell^{2} - a^{2}\right) - 2M\ell^{2}r_{+} + a^{2}\ell^{2} \approx r_{+}^{4} + \ell^{2}r_{+}^{2} - 2Mr_{+}\ell^{2} = 0,
\label{eq:horizon-equation}
\end{equation}

so that $r_{+} \approx \sqrt{2M\ell^{2}}^{1/2}$ to leading order.

Using this expansion, we derive the scaling relations quoted in \cref{sec:extensivity}:

\begin{align}
E &\sim r_{c}^{2} \left(\frac{r_{+}^{2}}{\ell^{2}}\right), \\
S &\sim r_{c}^{2} \left(\frac{r_{+}}{r_{c}}\right)^{2}, \\
\mathcal{P} &\sim \frac{r_{+}^{2}}{\ell^{2}}.
\end{align}
\label{eq:scaling-relations}

These scaling relations confirm that in the large black hole limit, the system behaves as a $(2+1)$-dimensional thermal fluid on the boundary.

\bibliographystyle{elsarticle-num}
\bibliography{references}

\end{document}